\documentclass[]{article}
\usepackage{url,multirow,morefloats,floatflt,cancel,tfrupee}

\usepackage{array,amsfonts,amssymb,amsbsy,latexsym,amsmath,tabularx,tabulary,graphicx,times,caption,fancyhdr,hyperref,placeins,subcaption,booktabs,lscape,float,threeparttable,titlesec}
\usepackage[utf8]{inputenc}

\makeatletter
\AtBeginDocument{\@ifpackageloaded{textcomp}{}{\usepackage{textcomp}}}
\makeatother
\usepackage{colortbl}
\usepackage{xcolor}
\usepackage{pifont}
\usepackage[nointegrals]{wasysym}
\usepackage{subcaption}
\makeatletter

\def\mcWidth#1{\csname TY@F#1\endcsname+\tabcolsep}

\def\cAlignHack{\rightskip\@flushglue\leftskip\@flushglue\parindent\z@\parfillskip\z@skip}
\def\rAlignHack{\rightskip\z@skip\leftskip\@flushglue \parindent\z@\parfillskip\z@skip}

\@ifundefined{etal}{}{}

\usepackage{ifxetex}
\ifxetex\else\if@twocolumn\@ifpackageloaded{stfloats}{}{\usepackage{dblfloatfix}}\fi\fi

\AtBeginDocument{
\expandafter\ifx\csname eqalign\endcsname\relax
\def\eqalign#1{\null\vcenter{\def\\{\cr}\openup\jot\m@th
  \ialign{\strut$\displaystyle{##}$\hfil&$\displaystyle{{}##}$\hfil
      \crcr#1\crcr}}\,}
\fi
}

\AtBeginDocument{%
  \@ifpackageloaded{endfloat}%
   {\renewcommand\efloat@iwrite[1]{\immediate\expandafter\protected@write\csname efloat@post#1\endcsname{}}}{\newif\ifefloat@tables}%
}%

\def\BreakURLText#1{\@tfor\brk@tempa:=#1\do{\brk@tempa\hskip0pt}}
\let\lt=<
\let\gt=>
\def\processVert{\ifmmode|\else\textbar\fi}

\@ifundefined{subparagraph}{
\def\subparagraph{\@startsection{paragraph}{5}{2\parindent}{0ex plus 0.1ex minus 0.1ex}%
{0ex}{\normalfont\small\itshape}}%
}{}

\newcommand\role[1]{\unskip}
\newcommand\aucollab[1]{\unskip}
  
\@ifundefined{tsGraphicsScaleX}{\gdef\tsGraphicsScaleX{1}}{}
\@ifundefined{tsGraphicsScaleY}{\gdef\tsGraphicsScaleY{.9}}{}
\def\checkGraphicsWidth{\ifdim\Gin@nat@width>\linewidth
	\tsGraphicsScaleX\linewidth\else\Gin@nat@width\fi}

\def\checkGraphicsHeight{\ifdim\Gin@nat@height>.9\textheight
	\tsGraphicsScaleY\textheight\else\Gin@nat@height\fi}

\def\fixFloatSize#1{}
\let\ts@includegraphics\includegraphics

\def\inlinegraphic[#1]#2{{\edef\@tempa{#1}\edef\baseline@shift{\ifx\@tempa\@empty0\else#1\fi}\edef\tempZ{\the\numexpr(\numexpr(\baseline@shift*\f@size/100))}\protect\raisebox{\tempZ pt}{\ts@includegraphics{#2}}}}

\AtBeginDocument{\def\includegraphics{\@ifnextchar[{\ts@includegraphics}{\ts@includegraphics[width=\checkGraphicsWidth,height=\checkGraphicsHeight,keepaspectratio]}}}

\DeclareMathAlphabet{\mathpzc}{OT1}{pzc}{m}{it}

\def\URL#1#2{\@ifundefined{href}{#2}{\href{#1}{#2}}}

\def\UrlOrds{\do\*\do\-\do\~\do\'\do\"\do\-}%
\g@addto@macro{\UrlBreaks}{\UrlOrds}

\edef\fntEncoding{\f@encoding}

\makeatother

\newif\ifmultipleabstract\multipleabstractfalse%
\makeatletter

\def\wileyIndent{1pt}
\usepackage[paperheight=10in,paperwidth=6.5in,margin=2cm,headsep=.5cm,top=2.5cm,headheight=1cm]{geometry}

\renewenvironment{abstract}
{\vspace*{-1pc}\trivlist\item[]\leftskip\wileyIndent\hrulefill\par\vskip4pt\noindent\textbf{\abstractname}\mbox{\null}\\}{\par\noindent\hrulefill\endtrivlist}

\usepackage[]{footmisc}

\def\author#1{\gdef\@author{\hskip-\dimexpr(\tabcolsep)\hskip\wileyIndent\parbox{\dimexpr\textwidth-\wileyIndent}{\centering\bfseries#1}}}

\def\title#1{\linespread{1}\gdef\@title{\centering\bfseries\ifx\@articleType\@empty\else\@articleType\\\fi#1}}

\let\@articleType\@empty \def\articletype#1{\gdef\@articleType{{\normalfont\itshape#1}}}

 \def\audegree#1{}

\newcolumntype{L}[1]{>{\raggedright\arraybackslash}p{#1}}

\date{}

\makeatother

\usepackage[T1]{fontenc}
\makeatother

\def\thanksspace{{\phantom{\textsuperscript{\thefootnote}}}}

\begin{document}

\title{RetailSynth: Synthetic Data Generation for Retail AI Systems Evaluation}

\author{
    Yu~Xia\thanks{Yu.Xia@Bain.com}{\thanksspace},\space Ali~Arian\thanks{Ali.Arian@Bain.com}{\thanksspace},\space Sriram~Narayanamoorthy\footnote{Corresponding author: Sriram.Narayanamoorthy@Bain.com},\space and Joshua~Mabry\footnote{Corresponding author: Joshua.Mabry@Bain.com}
    \normalsize\normalfont\itshape \\Bain \& Company, Inc.
}

\maketitle
\begin{abstract}

Significant research effort has been devoted in recent years to developing personalized pricing, promotions, and product recommendation algorithms that can leverage rich customer data to learn and earn. Systematic benchmarking and evaluation of these causal learning systems remains a critical challenge, due to the lack of suitable datasets and simulation environments. In this work, we propose a multi-stage model for simulating customer shopping behavior that captures important sources of heterogeneity, including price sensitivity and past experiences. We embedded this model into a working simulation environment --- RetailSynth. RetailSynth was carefully calibrated on publicly available grocery data to create realistic synthetic shopping transactions. Multiple pricing policies were implemented within the simulator and analyzed for impact on revenue, category penetration, and customer retention. Applied researchers can use RetailSynth to validate causal demand models for multi-category retail and to incorporate realistic price sensitivity into emerging benchmarking suites for personalized pricing, promotions, and product recommendations. 

\def\keywordstitle{Keywords}

\end{abstract}
\section{Introduction}

With the growth of digital marketing and e-commerce, retailers have invested in building AI systems for targeted sales promotions, dynamic pricing in brick-and-mortar and online stores, product search and recommendation services, and online advertising \cite{katsov_introduction_2017}. In diverse fields such as econometrics, marketing science, recommendation systems, and reinforcement learning (RL), there has been relentless development of ever-more sophisticated predictive models and decision-making algorithms for retail marketing. Progress is often claimed if a new model is better in predicting held-out data than previous ones or a new algorithm performs better in an isolated A/B test, but reliability and reproducibility challenges are well-documented \cite{cremonesi_progress_2021}. Consequently, there has been a shift towards building data-centric AI platforms that host benchmark datasets, online evaluation methods, and baseline implementations \cite{mazumder_dataperf_2023}.

Nonetheless, the challenge of reliably evaluating and benchmarking retail AI systems is growing. Practitioners are shifting focus from simple item recommendation to value-aware recommendation and causal decision-making algorithms that are designed to dynamically optimize the economic value from recommendations for both firms and consumers \cite{fernandez-loria_causal_2022}. However, evaluation of these more complex systems requires more complete, longitudinal datasets. Existing public datasets are unreasonably small when compared to typical industry datasets, collected under biased and often unknown marketing policies, and missing many key fields required to accurately model customer behavior \cite{de_biasio_systematic_2023}. Within the retail industry, there are privacy and competitive concerns that make it unlikely the status quo of poor availability of realistic public datasets will change anytime soon. In addition, there are business constraints and fairness concerns that prevent aggressive experimentation across the marketing mix---product, place, promotion, and price. Within the AI community, offline evaluation based on static datasets remains the de facto standard, due to its simplicity and low computational cost. However, it will remain difficult to achieve a mechanistic understanding of AI systems, relying only on offline evaluation of historical data. 

Given these challenges, we expect an increasing reliance on Monte Carlo simulation for assessing short-term and long-term impacts of AI systems and for assessing their performance against strategic objectives. More specifically, we see a need for access to fully synthetic data that reasonably mimics customer buying behavior to facilitate collaboration between academic and industrial researchers, without risking customer privacy or trade secrets and to test the robustness of AI systems to changes in the environment \cite{ie_recsim_2019, broderick_toward_2023}. 

To enable building realistic evaluation environments for retail AI, we introduce the \textbf{RetailSynth} simulation environment, focusing initially on customer-level simulation of multi-category retail, with price as the primary decision variable \footnote{Code available at \url{https://github.com/RetailMarketingAI/retailsynth}}. Our simulation environment addresses several key challenges specific to retail marketing:
\begin{enumerate}
  \item Extending econometric-style generative models to cover the full customer life-cycle. 
  \item Creating realistic differences in price sensitivity across customers and products.
  \item Generating synthetic customer trajectories for large numbers of products and customers efficiently. 
\end{enumerate}

We develop an interpretable multi-stage decision framework that leverages econometric theory and covers the full set of decisions: visiting a store, selecting categories and products to purchase, and deciding the quantity to purchase. Each decision stage is sensitive to individual customer preferences, as represented by the utility values for products within the retailer's assortment. Product utilities are determined by the union of customer and product attributes and marketing mix variables; keeping product utilities central to our framework makes it both interpretable and simple to calibrate to real-world conditions. We provide a detailed account of how we calibrated the synthetic data model and show how the choice distributions and aggregate buying behavior compare to a real-world dataset. Following econometric theory, we calculated the elasticity for each decision-stage and used these results to verify that the sensitivity of consumers to price is heterogeneous and leads to realistic differences across customer segments. In addition to verifying the simulator agrees with the theory at the detailed level, we present a detailed scenario analysis, showing how customer demand shifts under realistic baseline pricing policies that vary the depth and frequency of promotional discounts. We show that, in addition to capturing broad behavioral changes in response to discounted prices, the calibrated simulator produces complex, heterogeneous behavior across customer segments differing in their price sensitivity. This differential price response across customers implies that AI-driven personalization can result in a significant uplift in our environment and motivates future applications of the simulator.

Our ambition is to build a full ecosystem of evaluation tools for retail AI. RetailSynth can serve immediately as a fundamental building block to test algorithms that aggressively flex marketing-mix variables, providing a simplified, ground-truth model of customer behavior that we expect to be applicable across domains ranging from grocery to fashion. As designed, RetailSynth can scale to predict shopping behavior across product assortments and customer populations of arbitrary size and enables rigorous evaluation of complex predictive models and recommendation algorithms through simulation. As data-sharing and experimentation become more common in the retail ecosystem, we expect data-driven methodologies for synthetic data generation to become more common and more useful. Accordingly, we provide a detailed account of the calibration procedure and our methodology for evaluating the fidelity of synthetic data to the real-world retail dataset so that others can follow the same playbook for validating future synthetic data generation schemes. We expect to see retail AI systems take over strategic decisioning from human managers only once the safety, fairness, and reliability of these systems has been well-established. It is our hope that this simulation framework can help accelerate the ongoing development and evaluation of retail AI systems.

\section{Related Works}

\subsection{Synthetic data generation}
The value of synthetic data for accelerating the development of AI systems has been broadly recognized, with a recent emphasis on building generative models of the data in its raw tabular forms, rather than modeling features derived from aggregated and transformed data \cite{patki_synthetic_2016, jordon_synthetic_2022}. Privacy-preserving modeling techniques (such as Generative Adversarial Networks, Variational Autoencoders, and Bayesian Networks) have been applied to generate synthetic data, while conforming to differential privacy guarantees \cite{wilde_foundations_2021}. For example, Athey and Imbens \cite{athey_estimating_2020} leveraged Wasserstein Generative Adversarial Networks to generate artificial data for evaluation of causal effects estimators. These privacy-preserving modeling techniques are powerful when sufficient historical data is available to learn an accurate data-generating process. When data is scarce, there is a rich tradition within econometrics of specifying structural causal models for generating synthetic data that can be used to validate new models and algorithms, with the caveat that the data may be highly stylized and lack the complexities of real-world data \cite{parikh_validating_2022}. For example, Andrews and co-workers \cite{andrews_comparison_2011} simulated customer brand-choice within a single category based on a nested logit model---an approach which we extend in this work to encompass multiple categories. 

\subsection{Consumer choice modeling}
To generate synthetic data, we rely on models that fit in the long tradition of consumer choice modeling based on utility-maximization theory \cite{fishburn_utility_1968}. In retail marketing, these models are typically used to inform marketing-mix decisions and resource allocation; over the past several decades, there has been recognition that jointly modeling purchase incidence, brand choice, and purchase quantity makes the models more accurate and is required to fully capture consumers' decision-making process \cite{chiang_simultaneous_1991}. Extending this approach to big data, McAuley and team developed a nested, feature-based matrix factorization framework to jointly model preferences and price sensitivities of individual consumers and showed that this approach could scale across large, multi-category assortments \cite{wan_modeling_2017}. They modeled each decision stage of category choice, product choice, and item quantity selection independently and showed that the model could accurately predict future consumer behavior. They did not, however, consider the impact of changes in promotion policy (i.e. discount depth and frequency) on consumer purchasing behavior, and so it remains untested whether such a complex model can be reliably used for personalized promotion optimization and yield increases in customer lifetime value. We built RetailSynth to help answer these questions about how consumers respond to large changes in price and promotion policy, and we compare its capabilities to a variety of other methods in the Discussion section \ref{discussion}.

\subsection{Customer Lifetime Value}
Understanding consumer behavior over longer time periods is typically accomplished with Customer Lifetime Value (CLV) models and is considered challenging due to the need to predict future purchases based on limited purchase history and dynamic preferences. When using CLV models to plan promotions, customers are typically ranked by expected value to the firm and different marketing interventions are applied to different customers \cite{ekinci_using_2014}. Common CLV models assume that customer purchase behavior is defined by frequency and average transaction value and that customers become inactive after a limited period \cite{romero_partially_2013}. One of the earliest models to take hold for predicting time between purchase events was the negative binomial distribution, which relies on the simple assumption that customers have individual purchase rates that follow a Poisson distribution \cite{fader_counting_2005}. In the big data era, it makes more sense, however, to use approaches, such as Hidden Markov Models (HMMs) or deep-learning-based sequence models, that account for customer features and marketing activities \cite{fader_probability_2009,morrison_generalizing_1988,montoya_dynamic_nodate,netzer_hidden_2008}. In that spirit, we incorporate customer features and marketing mix variables into our model for customer purchase events and store visit behavior so that the customer lifetime value estimate is dynamic and responsive to changes in marketing policy. 

\subsection{Evaluation of AI systems}
Cutting-edge methods integrate customer value measurement with online decisioning. For example, Liberali and Ferecatu blended bandit decision-making with dynamic customer value estimation using HMMs; however, initialization was costly, requiring a randomized controlled trial to generate unbiased model calibration data and avoid a "cold start" \cite{liberali_morphing_2022}. For recommendation systems, the cold-start problem is well-known and one of the key reasons that simulation is required for assessing the dynamic performance of systems that must learn and earn \cite{gauci_horizon_2019, hafner_tensorflow_2018, santana_mars-gym_2020, ie_recsim_2019}. The base components of these simulation systems are the environment and recommendation agent. The value estimate in the environment is typically based either on a causal model \cite{ie_recsim_2019} or off-policy evaluation of historical data, using methods like doubly robust evaluation \cite{santana_mars-gym_2020}. There is an immense variety of approaches to recommendation, ranging from static filtering to off-policy optimization to on-policy optimization, with RL approaches receiving much attention due to their ability to optimize for long-term rewards \cite{powell_reinforcement_2022}. Within retail, RL algorithms have been developed for dynamic pricing, \cite{kastius_dynamic_2022, rana_real-time_2014} and there have been early efforts to develop marketplace simulation platforms for algorithm evaluation, such as \textit{Price Wars} \cite{serth_interactive_2017}. We are not aware of simulation platforms like RetailSynth that link pricing and promotions policies to individual customer decision-making across the full lifecycle from visiting a store to purchasing an individual item.
\section{Multi-stage data generating framework}\label{section: overall_framework}

\begin{figure}
	\centering
        \includegraphics[width=1\linewidth]{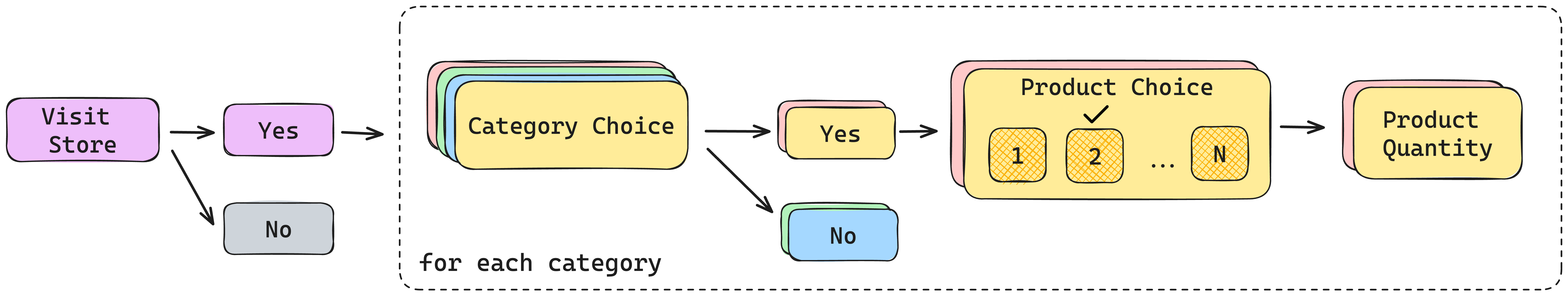}
	\caption{Generative model design and linkages between stages of the customer decision process in RetailSynth.}
	\label{fig:fig1}
\end{figure}

The purchasing behavior of retail customers represents a nuanced journey that is influenced by a number of factors. For example, a shopping event could originate from a spontaneous impulse, a marketing trigger, or a need for item replenishment. Furthermore, the decision to purchase a specific product may involve careful evaluation of alternatives and consideration of price, brand, and product quality. This study introduces a novel and interpretable multi-stage retail data synthesizer aimed at capturing the complex nature of the customer decision-making process. We propose a factorized model that consists of four key decision stages (Figure \ref{fig:fig1} provides a visual representation of the basic synthesizer model design):
\begin{itemize}
\item Whether to visit the store (Store visit choice)
\item Whether to make a purchase within a category (Category choice)
\item Which product to buy from a category (Product choice)
\item How much of the product to buy (Purchase quantity)
\end{itemize}

This model is designed to generate a record of customer purchases across an entire firm over a fixed time period for a finite product assortment without inventory constraints. We let $u$ be the customer index $u = (1, 2, \ldots, U)$, $t$ be the time index $t = (1, 2, \ldots, T)$, $i$ be the product index $i = (1, 2, \ldots, I)$, and $j$ be the category index $j = (1, 2, \ldots, J)$. The total number of customers is represented by $N_u$, products by $N_i$, and categories by $N_j$. One product can belong to only one of the $N_j$ categories. We represent this many-to-one relationship between products and categories using a function $f: I \to J$. This function maps each product $i$ to its corresponding category $j$:
\[f(i) = j, \quad i \in I, \, j \in J.\]

We can now define the probability of customer $u$ purchasing $Q$ quantity of product $i$ using the conditional probability (we develop the structure for each time step and so do not use an index for time step in the following presentation for simplicity):
\begin{align*}
&P(Q_{ui}=q,S_u,C_{uj},B_{ui})\\
&=P(S_u)P(C_{uj} | S_u) P(B_{ui}|S_u,C_{uj})P(Q_{ui}=q | S_u,C_{uj},B_{ui}) 
\end{align*}

where:
\begin{align*}
S_u &= \text{binary outcome (visited versus did not visit store) for customer $u$} \\
C_{uj} &= \text{binary outcome (category purchased versus did not purchase) } \\
          &\text{\quad$\>$ for customer $u$ and category $j$}\\
B_{ui} &= \text{binary outcome (product purchased versus did not purchase)} \\
          &\text{\quad$\>$ for customer $u$ and product $i$}\\
Q_{ui} &= \text{units purchased by customer $u$ of product $i$}\\
\end{align*}

Each component of the above can be encoded using a generative model. We follow a hierarchical modeling approach similar to \cite{andrews_comparison_2011, wan_modeling_2017, donnelly_counterfactual_2019} to specify the various sub-models, while taking note of several key considerations. First, the decision-making process is not always linear and may involve multiple iterations across these steps --- we make a simplifying assumption that each customer purchase decision is conditionally independent. Second, our model adopts a simplifying assumption of independence between category choices. This assumption is essential to avoid the curse of dimensionality, which might make the simulation intractable at industrial scale. However, in reality, individual category purchase decisions are interdependent, often influenced by factors such as product complementarity. Third, in real-world systems, individual decision-makers show significant variation in product preferences and price sensitivity due to unobserved and idiosyncratic factors that may vary over time. While we incorporate latent heterogeneity into the data synthesizer, we do not address time-varying changes in customer preference.

\subsection{Customer product utility}

Embedded in all of the choice decisions enumerated here is the notion of customer utility. The random utility for product alternative $i$ for customer $u$ at time $t$ takes the form
\begin{align}\label{eq:random_utility_product}
    \mu^{prod}_{uit} &= \mathbf{\beta_{ui}^x} \mathbf{X_{uit}} + \mathbf{\beta_{ui}^w} \mathbf{W_{uit}} + \epsilon_{uit}
\end{align}

The utility term is composed of four components:
\begin{itemize}
\item $\mathbf{X_{uit}}$ represents time-varying features that are customer-specific (e.g. category promotional offers based on loyalty status) or product-specific (e.g. product cost, brand equity).
\item $\mathbf{W_{uit}}$ represents treatment or decision variables that can be optimized (e.g. the primary marketing-mix variables of price, promotion, and place).
\item $\epsilon_{uit}$ is a noisy intercept used to capture the idiosyncratic nature of customer demand and follows an extreme value distribution with fat-tail effects.
\end{itemize}

In real-world settings, features in $\mathbf{X_{uit}}$ will often be confounded with the utility and with the treatment variables. For example, the brand strength of a product will influence its price, and both brand strength and price can affect the customer utility of a product. This formulation of the product utility allows us to flexibly incorporate policy variables into the $\mathbf{W_{uit}}$ term. As an example, if we want to use the synthesizer for analysis of pricing policies, where prices are set at the product level for each time step, we can simply express Equation \ref{eq:random_utility_product} as follows: 
\begin{align}\label{eq:product_utility}
    \mu^{prod}_{uit} &= \mathbf{\beta_{ui}^x} \mathbf{X_{uit}} + \beta_{ui}^w log(P_{it}) + \epsilon_{uit}
\end{align}

We show a more detailed example of one such pricing policy in our synthesizer in Section \ref{price_policy_section}. In the following sub-sections, we describe the data-generating process for each of the choice stages. For the remainder of this paper, we employ the following key notation: $\mathbb{I}_{ut}$ will indicate whether customer $u$ visits a store in week $t$, $\mathbb{I}_{ujt}$ will indicate whether customer $u$ initiates a purchase in category $j$ at week $t$, and $\mathbb{I}_{uit}$ will indicate whether the customer buys product $i$ at week $t$. Also, $J_j$ will denote the set of products in a specific category $j$.

\subsection{Product choice and product purchase quantity}
Product choice and purchase quantity decisions are expressed as a function of the product utility specified in Equation \ref{eq:random_utility_product}. To simplify the modeling of the product choice decision, we assume that once a customer has initiated a purchase within a category, they make a multinomial choice and buy exactly one product from the category. McAuley and co-workers \cite{wan_modeling_2017} give the example of purchasing milk, where the customer chooses a category like organic milk and then chooses between brands. Our multinomial logit choice model is given by 
\begin{align}\label{eq:product_choice}
    p^{prod}_{uit} &= P(\mathbb{I}_{uit} = 1 | \mathbb{I}_{ujt} = 1)\\
    &= \frac{\exp{(\mu^{prod}_{uit})}}{\sum_{k \in J_j} \exp{(\mu^{prod}_{ukt}})}
\end{align}

It is important to emphasize that since the product choice is conditional on the category choice, only the set of products within the chosen category enters the consideration set of the customer.

For a realized product choice $i$ for customer $u$ at week $t$, we use a shifted Poisson distribution to generate the quantity $y_{uit}$ purchased. The shifting of the distribution is necessary as we need to ensure that the customer purchases at least one unit of the chosen product based on the assumptions described in the previous step. The rate parameter, $\lambda_{uit}$, of the Poisson distribution is expressed as a function of the product utility as follows.
\begin{align}\label{eq:product_demand_rate_parameter}
    \lambda_{uit} &= \exp(\gamma^{prod}_{0i} + \gamma^{prod}_{ui}\ \mu^{prod}_{uit})
\end{align}

The $\gamma^{prod}_{0i}$ and $\gamma^{prod}_{ui}$ parameters can be used to scale the utility term to calibrate the conditional quantity purchased distribution. The distribution follows the standard form:
\begin{align}\label{eq:product_demand}
    P(Q_{uit} = q | \mathbb{I}_{uit}=1) &= \lambda_{uit} ^{q-1}\frac{exp(-\lambda_{uit})}{(q-1)!}
\end{align}

\subsection{Category choice}\label{sec:category_choice_stage}
For the category choice stage, we assume that a customer makes independent binary decisions on the purchase of products within each category, such that each decision is a Bernoulli trial. This assumption of category independence helps us scale the data synthesizer to real-world settings, as each category can be run in parallel. The assumption is likely to be violated if budgetary pressure is high or if there are products across categories that are highly complementary or substitutable \cite{donnelly_counterfactual_2019}. 

To quantify the customer's affinity for a specific category $j$ at time $t$, we rely on a category preference score, $CV_{ujt}$.
\begin{align}\label{eq:category_preference_score}
    CV_{ujt} &= \log \sum_{k \in J_j} \exp(\mu^{prod}_{ukt})
\end{align}
The category affinity score represents the maximum utility available from making a purchase within the category. The probability $p^{cate}_{ujt}$ of customer $u$ initiating a purchase of a product in category $j$ at time $t$ takes the form:
\begin{align}\label{eq:category_choice}
    p^{cate}_{ujt} &= P(\mathbb{I}_{ujt}=1 | \mathbb{I}_{ut}=1) \\
    &= \frac{\exp(\gamma_{0j}^{cate} + \gamma_{1j}^{cate} CV_{ujt})}{1 + \exp(\gamma_{0j}^{cate} + \gamma_{1j}^{cate} CV_{ujt})}
\end{align}

Modifying the values of $\gamma_{0j}^{cate}$ and $\gamma_{1j}^{cate}$ parameters let us determine the relative importance of the category preference score on the category choice decision. 

\subsection{Store visit choice}
The store visit choice is the first choice a consumer must make in a given time period in our framework. Similar to category choice, we craft a store visit preference score, $SV_{ut}$, to capture the customer's future propensity to visit the store. This score represents the inclusive value of making a purchase from any of the categories in the store. This metric is defined as follows:
\begin{align*}
    SV_{ut} &= \log \sum_{j \in J} \exp(CV_{ujt})
\end{align*}
In our framework, the store visit is a binary choice to shop at any store or digital property belonging to the retailer, as we are not explicitly modeling individual stores. Therefore we employ a Bernoulli distribution to generate the probabilities. We assume that previous store visits and current marketing activity influence the decision in the current time period. In the absence of a store visit, the propensity to visit the store in the future decays. We formulate our model as follows: (1) The store utility is additive and depends on observable customer attributes and marketing activities, which are both encoded in $\mathbf{X^{store}_{ut}}$, and on the effect of previous store visits. Specifically, the store visit preference score from the previous time step, $SV_{u(t-1)}$ is added to the store utility if and only if the customer visited the store in that previous time step. This term accounts for the beneficial effect of repeat shopping on customer retention. (2) The customer probability to visit the store depends on the current propensity to visit the store, weighted by a factor of $(1-\theta_{u})$, and the store visit probability at the previous time step, weighted by a factor of $\theta_{u}$. The intention here is to balance short-term and long-term effects on customer retention. The process described here follows these equations:
\begin{align}
P(\mathbb{I}_{ut}) &=
    \begin{cases}
        1 & \text{if } t = 1, \\
        (1-\theta_{u}) s_{ut} + P(\mathbb{I}_{u(t-1)}) \theta_{u} & \text{if } t > 1,
    \end{cases} \label{eq:main_equation} \\
    \intertext{where:}
    \mu^{store}_{ut} &= \gamma_0^{store} + \mathbb{I}_{u(t-1)} \gamma_1^{store} SV_{u(t-1)} + \boldsymbol{\gamma_2^{store}} \mathbf{X_{ut}^{store}} \label{eq:mu_store}\\
    s_{ut} &= \frac{exp(\mu^{store}_{ut})}{ 1 + exp(\mu^{store}_{ut})} \label{eq:s_ut}
\end{align}

The $\gamma_{0}^{store}$, $\gamma_1^{store}$ and $\boldsymbol{\gamma_2^{store}}$ parameters can be tuned to shift and scale up the individual factors affecting the store utility thereby influencing the probability of a store visit.

\subsection{Price policy setting}\label{price_policy_section}

Typical retail datasets do not include a complete price history for all products, as they are produced by legacy systems that log only transactions. Therefore, we need to impute the complete price histories. A common approach is to assume the price at any time equals the most recently observed purchase price, but this approach can suffer from significant observational bias, if products are rarely purchased at their original price \cite{wan_modeling_2017}.

We adopt a systematic method of generating prices that reflects real-world practice. We take into account the following constraints: (1) A product has a baseline price that is usually stable over extended periods of time; (2) depending on the product type, there should be either frequent or infrequent discounting events; (3) discount depth should be less than the profit margin for the category. While there are a variety of pricing policies that fit within these constraints, we implement a high-low pricing strategy  as a reasonable baseline. 

To implement the high-low policy, we assume that the price for each product follows a first-order Markov process, where at each time step $t$, the product price is set to the base price (state $s =0$) or a discounted price (state $s =1$). There is then a stochastic transition to either state in the following time step. We initialize the system such that no product is in the discount state. The probability that the system will be in state $j$ given it was in state $i$ at the previous time step is given by $a_{ij} = p_{s(t+1)| s(t)}(j|i)$. Our two-state system has the following transition matrix: 
\begin{align}\label{eq:transmat}
  \boldsymbol{A} = \begin{bmatrix}
       1 - a_{01} & a_{01}  \\
       1-a_{11} & a_{11} 
  \end{bmatrix}
\end{align}

The transition probabilities are generated according to 
\begin{align}\label{eq:transition_probs}
    a_{ij} \sim  Beta(\alpha_{ij}^{trans},\beta_{ij}^{trans})
\end{align}

The HMM produces observations of the discount depth for product $i$, as given by 
\begin{align}\label{eq:product_price_hmm}
    D_{it} &= 
    \begin{cases} 
    0  & s=0\\ 
        Beta(\alpha^{disc},\beta^{disc}) & s=1\\ 
    \end{cases}
\end{align}

For each product, the price is given by
\begin{align}\label{eq:product_price}
    P_{it} &= 
    \begin{cases} 
    \alpha_{i0} + \alpha_1 Z_{i}  & s_d=0\\ 
        (1 - D_{it})(\alpha_{i0} + \alpha_1 Z_{i}) & s_d=1\\ 
    \end{cases}
\end{align}

where $\alpha_{j0}$ is an observed factor that affects product price, such as product cost, and $Z_{i} $ is an unobserved factor, such as brand strength, that affects both prices and demand, creating an endogeneity condition, and is modulated by the global parameter $\alpha_1$.

Based on this price generation process, the resultant product utility from Equation \ref{eq:product_utility} can be expressed as follows:
\begin{align}\label{eq:product_utility_w_endog}
    \mu^{prod}_{uit} &= \mathbf{\beta_{ui}^x} \mathbf{X_{uit}} + \beta_{ui}^{z} Z_{i} + \beta_{ui}^w log(P_{it}) + \epsilon_{uit}
\end{align}
We will use this as the specific form of product utility in the remainder of this study.

\subsection{Bayesian data generation process} \label{sec:model_spec}
\begin{figure}
	\centering
        \includegraphics[scale=0.375]{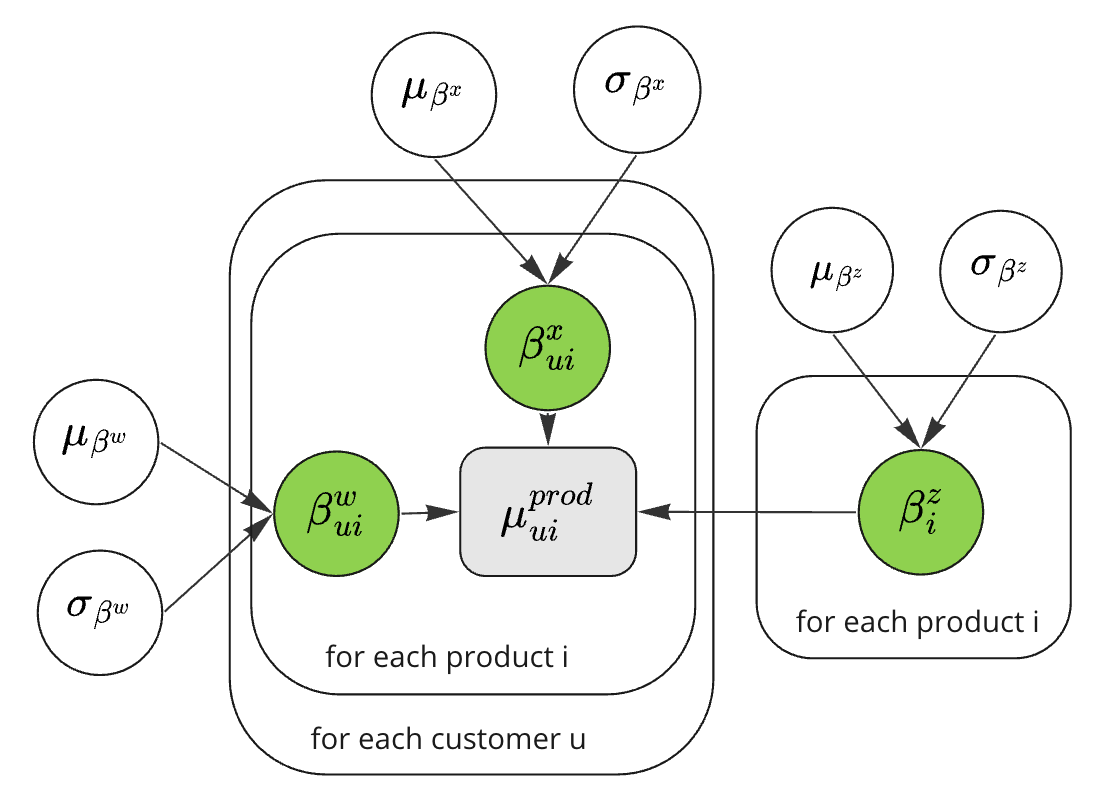}
        
\caption{Product utility model architecture using plate notation. White indicates model hyperparameters, green represents fixed effects, and grey is the utility outcome.}
	\label{fig:product_utility_plate}
\end{figure}

So far we have presented the functional form of the data generating functions. When generating sample data, we rely on a Bayesian framework, where all model parameters, random variables, and error distributions are drawn from fully-specified prior distributions. We give an example of a fully specified model in Figure \ref{fig:product_utility_plate}, which shows the product utility model in plate notation.  To generate a product utility per product and per customer, we need only specify the prior distributions for the parameters $\beta^w_{ui}$, $\beta^x_{ui}$ and $\beta^z_{i}$, as represented by $\mu_{\beta^w}$, $\sigma_{\beta^w}$, $\mu_{\beta^x}$, $\sigma_{\beta^x}$, $\mu_{\beta^z}$, and, $\sigma_{\beta^z}$ and then draw samples from those prior distributions for each customer and each product. We follow the same approach for each stage of the decision framework. The full set of priors and parameters used for the synthesizer is presented in Appendix \ref{appendix:parameter}.

In the following section, we describe how we calibrated the model hyper-parameters such that the distributions of metrics relevant to retail businesses are similar to a representative dataset. In addition to inspecting common metrics like customer recency, category penetration, basket size, and sales volume, we also inspect the product elasticity of demand. We present an analytical expression of the elasticity in the Appendix \ref{appendix:elasticity}.  Our formulation results in overall elasticity being additive, which allows us to compare the importance of price across the steps. We expect product choice elasticity to be the dominant effect based on prior work \cite{wan_modeling_2017}.

\section{Results}
We used a public dataset of grocery transactions to validate the assumptions made in our framework and to calibrate the synthesizer. In this section, we first provide a general overview of the dataset and a description of data cleaning. We next describe exploratory data analysis performed to help us understand the potential limitations of our proposed framework (and the dataset), with a focus on pricing, category choice, and consumer choice behavior. Following this analysis, we provide an overview of model calibration and discuss goodness of fit. We then describe how we used the calibrated synthesizer to simulate a range of pricing policies that are typical to retailers and explore the diverse resulting customer behavior, with a focus on how pricing affects customer retention, overall spending, and category penetration. 

\subsection{Data description and analysis}
Real-world data comes from the Dunnhumby - The Complete Journey \cite{dunnhumby_complete_2014} dataset. For ease of use, we accessed the data through the Python package \textit{completejourney-py} \footnote{https://pypi.org/project/completejourney-py/}. The dataset covers the purchasing behavior of more than two thousand households over a one-year period, who regularly visit the retailer, encompassing approximately 1.47 million transactions and featuring a diverse catalog of around 92,000 products. The product table included a comprehensive category hierarchy, covering product department, product category, and product type. Key transaction details, such as item quantity, transaction sales amount, and discounts, were recorded, enabling the derivation of unit prices and discounts. While household demographic data was available for a subset of 801 households, the lack of data for 60\% of the households led us to omit demographic information from our analysis. 

We conducted basic data cleansing by excluding transactions with non-positive quantities or sales amounts. Additionally, we de-duplicated the product metadata information and assigned unique identifiers to products and categories based on their hierarchical structure. This exercise reduced the unique number of products to $\sim$26,000. Our study's unit of analysis was at the customer-product-week level, and the data was aggregated accordingly. We introduced placeholder records for weeks when customers did not make a purchase, setting the quantity purchased to zero for those instances. As discussed in the Section \ref{price_policy_section}, due to how data is logged only for actual sales, the raw data captured only partial price history. In cases where no sales were observed for a product within a week, we set the product's price to the mean price across all available weeks.

\begin{figure}
    \centering
    \includegraphics[width=0.6\textwidth]{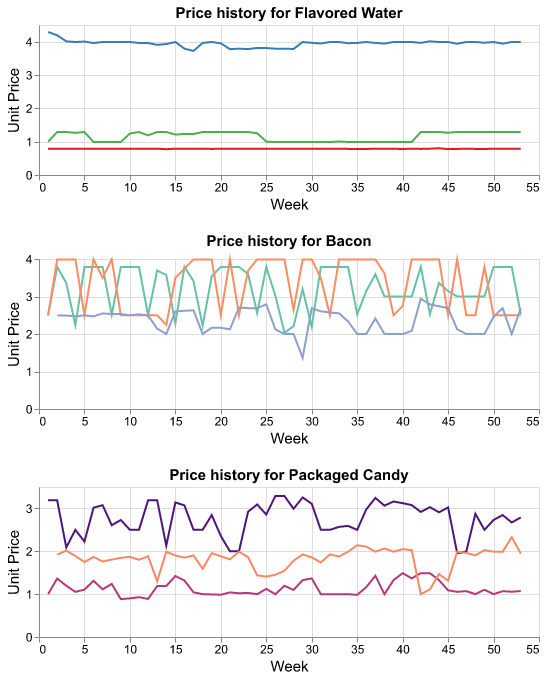}
    \caption{Historical price trends for selected products in the Flavored Drinks, Bacon, and Packaged Candy categories over successive weeks. Each product is shown in a different color.} 
    \label{fig:price_history_real_data}
\end{figure}

\subsubsection{Pricing data analysis}
We derived important insights into both the pricing strategy adopted by the retailer and household buying behavior through our analysis of purchase histories. Across all transactions, 50\% of transactions have a discount involved, at an average discount depth of 26\% with a standard deviation of 15\%. For most products, the number of transactions each week was not sufficient for us to infer the details of the pricing policy. For some more frequently purchased products, we looked at the historical pricing trends over successive weeks. Figure \ref{fig:price_history_real_data} illustrates the unit price trajectory for a randomly selected trio of high-volume products within each of the Flavored Drinks, Bacon, and Packaged Candy categories. From inspection of the data, any notion that product pricing was set randomly is immediately dispelled. Instead, there often appears to be a baseline price and discounts offered at varying discount depths and frequencies. There are products that were seldom discounted (e.g., the Flavored Water category) as well as products that were frequently discounted (e.g., the Bacon category, the Packaged Candy category). Within products that got a discount, there is a difference in discount depth and frequency. These observations support using the HMM pricing generator described above in Section \ref{price_policy_section}.

\subsubsection{Category choice analysis}\label{sec:category_choice_analysis}

\begin{figure}
    \centering
    \includegraphics[width=0.6\textwidth]{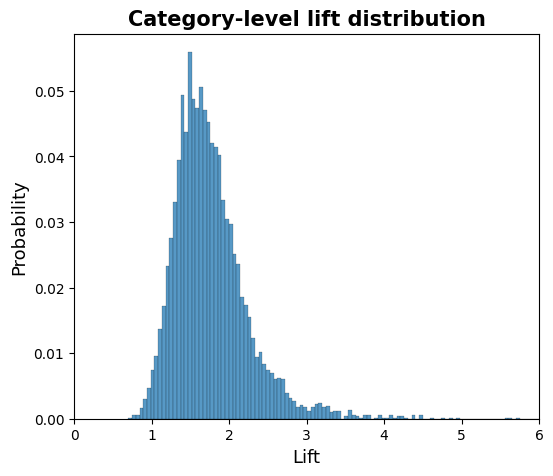}
    \caption{Distribution of lift values representing cross-category purchase associations.}
    \label{fig:apriori_real_data}
\end{figure}

We explored the customer purchasing behavior across categories, to determine how strongly the decision to purchase items from one category is influenced by the decision to purchase items from other categories. To accomplish this, we employed the Apriori Algorithm \cite{agrawal_fast_1996} at the category level to assess whether there is a significant tendency for categories to be bought together. Our analysis focused on pairs of categories, and we examined the lift values generated by the algorithm. If a lift value is much larger than 1, it suggests a positive dependence, i.e. a complementary effect. When a lift value is small, it suggests a negative dependence, i.e. a substitution effect. The distribution of lift values in Figure \ref{fig:apriori_real_data} shows that a significant number of category pairs exhibit a complementary effect. Notably, the $75^{th}$ percentile of the lift distribution is around 2, emphasizing that many category combinations exhibit complementary effects. In the categories that have particularly strong associations, there are common bundles such as pasta and pasta sauce, bread and butter, and dog food and pet care products. The Complete Journey data covers brick-and-mortar grocery stores, where it is likely that shopping trips are part of a larger weekly planning and replenishment cycle that may further encourage purchasing of specific product bundles. When modeling an Instacart dataset, Patidar and co-workers \cite{tulabandhula_multi-purchase_2023} found that while ~70\% of product pairs had no significant association, there were many product pairs with strong complementarity and the complementary item most often belonged to the produce category, which makes sense in the context of meal planning.

Contrary to these empirical observations, our framework assumes that customers make independent decisions regarding the purchase or non-purchase of products within each category (see Section \ref{sec:category_choice_stage}). We feel that brick-and-mortar grocery shopping may be one domain that could benefit from a more complex purchase decision model than that presented here. The assumption of category choice independence may be more suitable for e-commerce platforms characterized by smaller baskets, specialty stores offering niche products (e.g., electronics), and businesses with infrequent consumer purchasing cycles (e.g., home improvement and furniture).  

\subsection{Synthesizer specification and calibration}

To calibrate the synthesizer, we set up a simulation of 100 customers over a period of 53 weeks utilizing a product catalog that closely mirrors the Complete Journey dataset, encompassing $\sim$ 26,000 products across 300 categories. 

This setup allowed for a comprehensive analysis of dynamic consumer preferences and behaviors across diverse product types. We configured the simulator such that we could manipulate prices and capture the effect on individual customer demand and customer retention. Specifically, we adopted the product utility formulation from Equation \ref{eq:product_utility}. While most of the terms in that equation are simple random variables, we assumed the customer-product price sensitivity $\beta_{ui}^{w}$ followed a multiplicative latent factor model, previously introduced in \cite{donnelly_counterfactual_2019}. 
\begin{align}
  \beta_{ui}^w &= c \beta_u^w \beta_i^w
\end{align}

Here, $\beta_{i}^{w}$ is a product-specific factor, $\beta_{u}^{w}$ is a customer-specific factor, and $c$ is a scaling factor to adjust the overall degree of sensitivity. Within the synthesizer, product price setting was based on the HMM outlined in Section \ref{price_policy_section}, with parameters set to reflect common practice within retail (Table \ref{tab:baseline_disc}). The transition probabilities were drawn from Beta distributions, with priors set so that the majority of the products were in the base price state, $s =0$, for the majority of the time, and discounting events, $s =1$, were short-lived. When a product was in a discount state, the discount magnitude $D_{it}$ was sampled from a Beta distribution, yielding an average effective discount of approximately 5.7\%. 

\FloatBarrier

\begin{table}[ht]
    \centering
    \begin{threeparttable}
        \footnotesize
        \caption{Discounting policy for model calibration applied to all customers and products. Discount state probability and expected discount depth are computed as described in Appendix \ref{appendix:scenario_def}.}
        \renewcommand{\arraystretch}{1.5}
        \begin{tabular}{p{1.2cm}p{1.2cm}p{1.2cm}ccc}
            \toprule
            \multirow{2}{1.2cm}{\textbf{Effective \newline discount}} &
            \multirow{2}{1.2cm}{\textbf{Discount state \newline probability}} & 
            \multirow{2}{1.2cm}{\textbf{Expected discount depth}}  & 
            \multirow{2}{*}{$\alpha_{01}^{trans}, \beta_{01}^{trans}$} & 
            \multirow{2}{*}{$\alpha_{11}^{trans}, \beta_{11}^{trans}$} &
            \multirow{2}{*}{$\alpha_d, \beta_d$} \\
            &&&&&\\
            \midrule
             5.7\% & 19\% & 30\% & (1, 5) & (2, 5) & (15, 35) \\
            \bottomrule
        \end{tabular}

        \label{tab:baseline_disc}
    \end{threeparttable}
\end{table}

In the remainder of this section, we discuss how we calibrated the synthesizer to match selected probability distributions from the Complete Journey data. Our calibration process was guided by two key objectives. The primary objective was to match the choice probability distributions from the four decision stages outlined in Section \ref{section: overall_framework} as closely as possible to the real-world data (see Figure \ref{fig:outcome_distribution_fit}). Our secondary objective was to ensure agreement of key customer behaviors such as time since last purchase (customer recency), total categories purchased per customer per week (category penetration), total items purchased per customer per week (basket size), and total items purchased per product per week (sales volume) (see Figure \ref{fig:key_metric_distribution_fit}). We iteratively adjusted the form of the sub-models and exact parameters of the prior distributions to roughly meet these objectives (For an example of how the sub-models are specified, see Section \ref{sec:model_spec}.)

\begin{figure}[ht]
	\centering
  \includegraphics[width=0.5\linewidth]{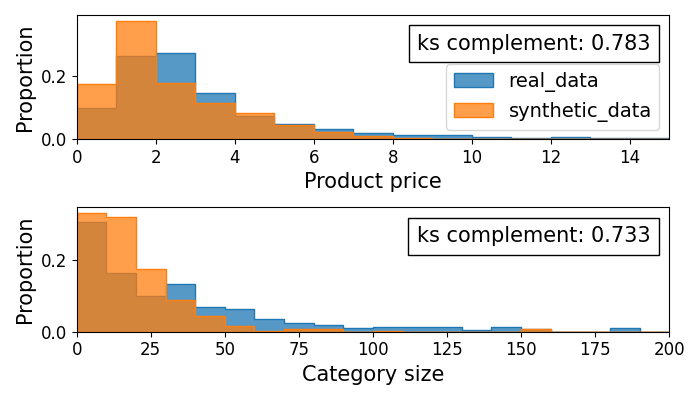}
	\caption{Observed price and category size distributions across all transactions in the synthetic and real (Complete Journey) purchase history data. (Top) Distribution of average price of purchased products inclusive of discounts. (Bottom) Distribution of number of products purchased per category (category size).} 
	\label{fig:product_price_and_size}
\end{figure}

\begin{figure}[ht]
 \centering
\begin{subfigure}[t]{0.48\textwidth}
  \centering
   \includegraphics[width=\textwidth]{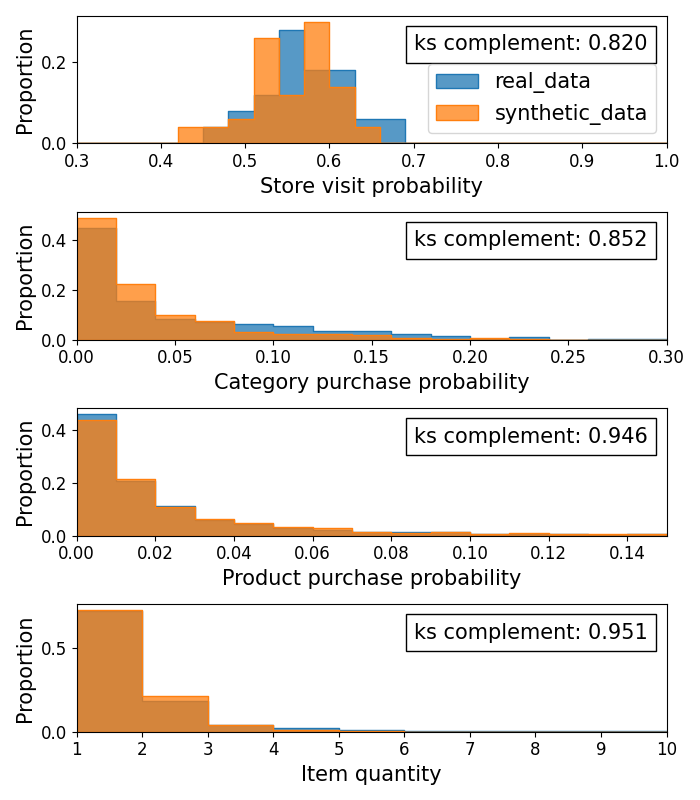}
   \caption{Step-wise choice probability distributions for the store visit, category choice, product choice, and purchase quantity decision stages observed in the synthetic and real (Complete Journey) purchase history data.}
  \label{fig:outcome_distribution_fit}
 \end{subfigure}
\hfill
 \begin{subfigure}[t]{0.48\textwidth}
  \centering
   \includegraphics[width=\textwidth]{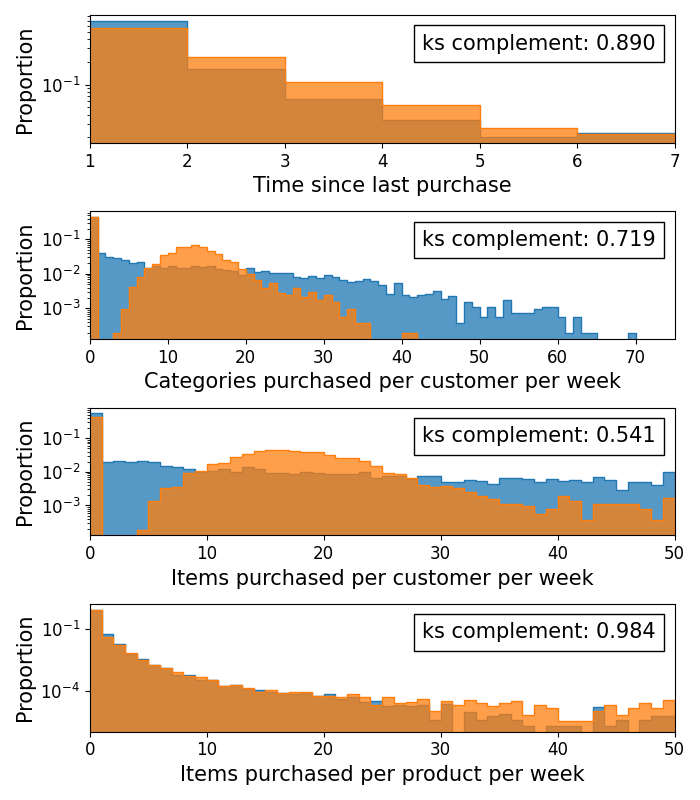}
   \caption{Aggregate metric distributions. From top to bottom: time since last purchase (customer recency), total categories purchased per customer per week (category penetration), total items purchased per customer per week (basket size), and total items purchased per product per week (sales volume) with y-axis in logarithmic scale.}.
  \label{fig:key_metric_distribution_fit}
 \end{subfigure}
 \caption{Synthesizer calibration results.} 
 \label{fig:calibration_results}
\end{figure}

Given the large number of hyper-parameters, we leveraged Bayesian optimization as implemented in Optuna \cite{akiba_optuna_2019}. We used the KS-Complement metric ($1 - \text{Kolmogorov-Smirnov statistic}$) to measure the accuracy of the data-generating process. This metric can take a value between 0 and 1 inclusive, where a score of 1 signifies identical distributions, while a score of 0 indicates entirely dissimilar distributions. Since we aimed to match multiple distributions, we calculated the KS-Complement metric for each distribution and set the optimization objective as a sum of the metric values computed for individual distributions. Our initial tuning runs focused on matching the category size and pricing distribution observed in the Complete Journey data. Figure \ref{fig:product_price_and_size} shows that the distributions fit the peaks of the distributions well, with KS-Complement > 0.7. However, the empirical distributions have heavier right tails than the synthetic data (Figure \ref{fig:product_price_and_size}). 

Through this initial tuning process, we discovered that we could not perfectly fit the heavy tails of many of the distributions due to the model assumptions and linkages between different choice stages. We observed that including the metric for every distribution of interest in our optimization scheme was counterproductive and often led to worse results than focusing on fitting the distributions that were more difficult to match. Consequently, our final tuning run maximized the metrics only for the category purchase probability and category count distributions.

This calibration strategy gave us fits that are not perfect but are more than suitable for the task at hand of building a simulator that can be used to understand and to compare various AI algorithms. More importantly, the customer behavior in the simulator is dynamic and heterogeneous, which we explore further in Section \ref{sec:scenarios}. For all customer decision stages, the KS-Complement metric in Figure \ref{fig:outcome_distribution_fit} consistently exceeds 0.82, highlighting a strong degree of similarity between the synthesizer's output and the real-world Complete Journey data. The fit is particularly impressive for the product purchase and item quantity distributions. The category choice probability, however, diverges with the synthetic data by having an excess of categories with low purchase probability and an extended tail. This mismatch, particularly at both distributional extremes, diminishes the accuracy of the aggregate purchase behavior. The basket size and category penetration distributions for the synthetic data (Figure \ref{fig:key_metric_distribution_fit}) have imperfect fits, which we attribute to basket composition. The synthetic data tends to include too few small baskets, containing 1-2 categories, and too few large baskets of 35 or more categories. This discrepancy is an expected side effect of not modeling product complementarity. As discussed in Section \ref{sec:category_choice_analysis}, we do not capture the sequential effect where customer purchases in one category lead to additional purchases in another and an overall larger basket with more categories purchased. Despite this shortcoming, the model accurately captures overall revenue and customer retention, as the distributions for customer recency and per-item sales volume show (Figure \ref{fig:key_metric_distribution_fit}). 

\begin{figure}[ht]
 \centering
\begin{subfigure}[t]{0.48\textwidth}
  \centering
   \includegraphics[width=\linewidth]{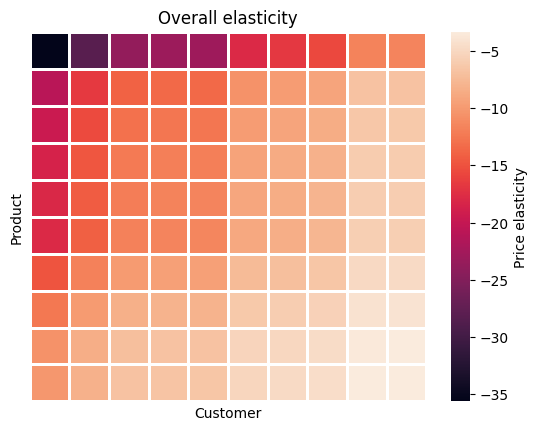}
   \caption{Price elasticities for randomly sampled products and customers in the synthetic dataset. Rows and columns sorted based on their average elasticity.}
   \label{fig:elasticity_heatmap}
 \end{subfigure}
 \hfill
 \begin{subfigure}[t]{0.48\textwidth}
  \centering
  \includegraphics[width=\linewidth]{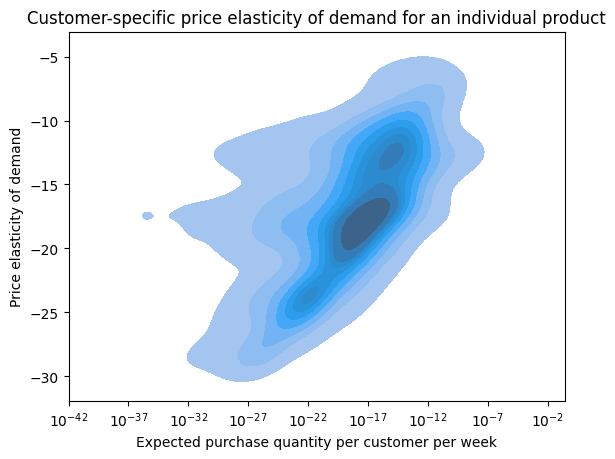}
   \caption{Price elasticity vs. expected purchase quantity of individual customers in the synthetic dataset. Data shown for a randomly selected product. Dark blue indicates the region of high probability mass.}
  \label{fig:elasticity_kde}
 \end{subfigure}
 \caption{Price elasticity in synthetic data.}
 \label{fig:calibration_results_elasticity}
\end{figure}

We also expect that the model can capture the dynamic behavior of customer segments with different price sensitivities in response to price changes and that the strength of the price response will vary across products, based on how we parameterized $\beta_{ui}^w$ to have product- and customer-specific latent factors. As noted in Section \ref{section: overall_framework}, there are, however, several additional sources of variation that can potentially drown out the price sensitivity. To demonstrate that the variation in price sensitivity is significant, we computed the price elasticity of demand per customer, per product using Equation \ref{eq:overall_elasticity}. The heatmap shown in Figure \ref{fig:elasticity_heatmap} visually represents the resulting price elasticity for a random sample of customers and products and shows a broad range of elasticities. We also examined the relationship between product preference and price sensitivity to determine if we observed a strong correlation. Figure \ref{fig:elasticity_kde} shows that customers with heightened price sensitivity (most negative elasticity) are the customers expected to buy the least amount of a given product on average. In agreement with previous finding \cite{wan_modeling_2017}, a majority of customers fall within a relatively narrow range of price sensitivities and that the more price-sensitive customers display a wider range of expected purchase amounts. This observation is consistent with the intuition that customers with a strong preference for a product are less price-sensitive, whereas those with a weak preference have a range of price sensitivities. 

\subsection{Scenario analysis}
\label{sec:scenarios}
Having established that there was variability in customer response to prices, we wanted to validate that the magnitude of the response at the individual level was large enough to significantly affect the aggregate measure of a retail firm's performance under a range of realistic pricing scenarios. We also wanted to show that there would be a potential uplift if the pricing policies were personalized, as personalization is a widespread and valuable strategy in retail and a key driver of AI adoption.  

We designed an experiment where we systematically varied discount depth and frequency and then monitored the effect on customer spending, category penetration, and retention while accounting for the potential profit margin impact. Our primary hypothesis was that our metrics would become saturated under different minimum discount depths and frequencies for different customer segments, indicating there would be a potential benefit to deploying AI-powered personalization. We also systematically screened for an effect of discount frequency since low-frequency discount events avoid the risk of customer fatigue, while more frequent discounting can be expected to improve customer retention. We wanted to make sure that marketing resource allocation was reflected in the scenarios. We posited that at any given time, the intensity of discounting and the intensity of marketing should be highly correlated and that this should have a strong impact on consumer perception of the firm. Recall from Equation \ref{eq:mu_store} that marketing affects the store visit utility through the term $\mathbf{X_{ut}^{store}}$. We simplified this term to be equal to the magnitude of discounts across all products, which can be expressed as follows:
\begin{align}
    \mathbf{X_{ut}^{store}} = \sum_i{D_{it}}
    \label{eq:store_utility_scenario}
\end{align}

\begin{figure}
    \captionsetup[subfigure]{position=top}
    \raggedright  
    \begin{minipage}{0.55\textwidth}
        \begin{subfigure}{\textwidth}
        \caption{}
        \includegraphics[width=\linewidth]{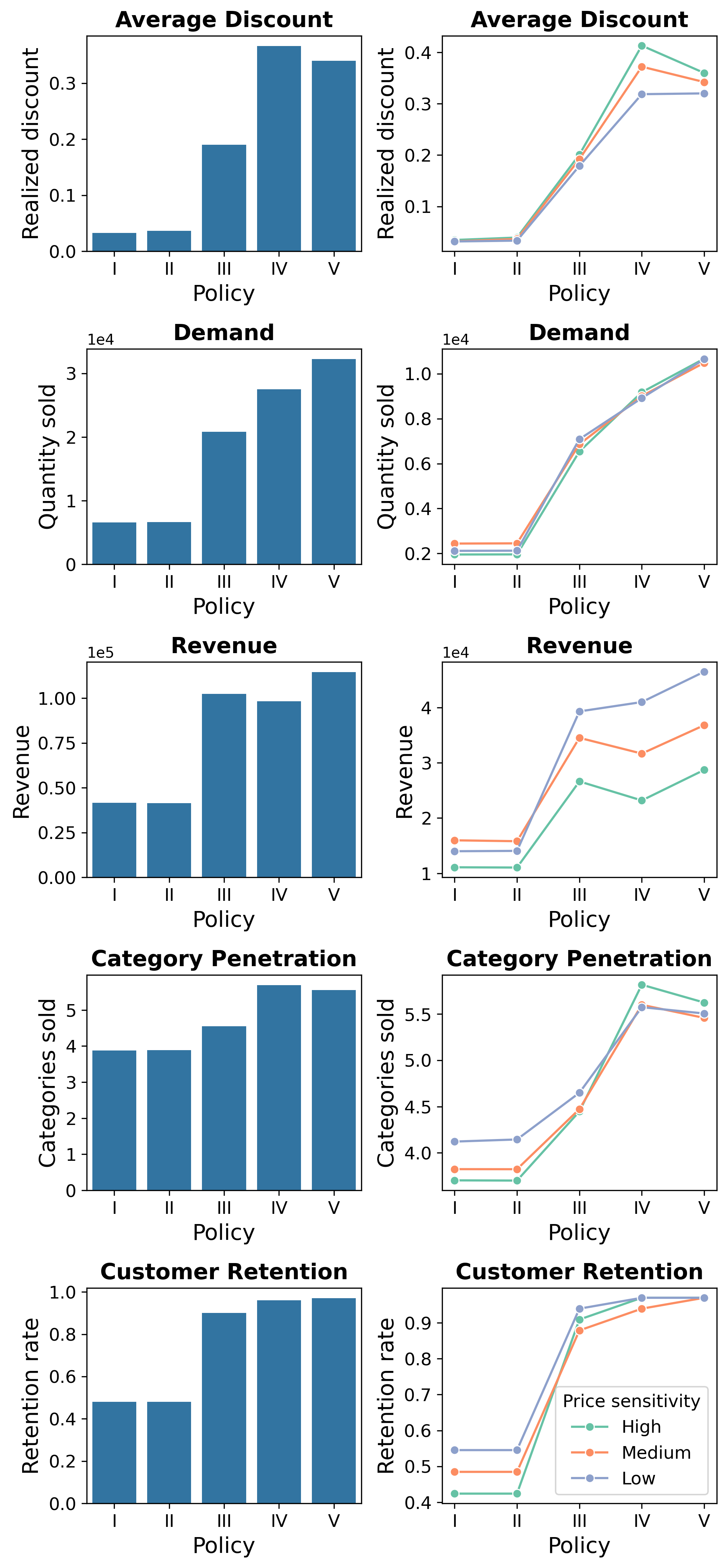}
        \label{fig:scenario_chart}
        \end{subfigure}
    \vspace{0.4cm}  
        \begin{subfigure}{\textwidth}
            \caption{}
            \footnotesize
            \renewcommand{\arraystretch}{1.5}
            \begin{tabular}{{cp{1.4cm}p{2cm}p{2cm}}}
                \toprule
                \textbf{Policy} & \textbf{Effective discount} & \textbf{Discount state \newline probability} & \textbf{Expected \newline discount depth}  \\
                \midrule
                 I & 3\%   & 60\% & 5\% \\ 
                 II & 3\%    & 30\% & 10\% \\ 
                 III & 15\%  & 60\% & 25\% \\
                 IV & 15\% & 30\% & 50\% \\
                 V & 24\%   & 60\% & 40\%\\
                \bottomrule
            \end{tabular}
            \label{tab:scenario_disc}
        \end{subfigure}
    \end{minipage}
    \hfill
    \begin{minipage}{0.3\textwidth}
    \caption{Effect of discount frequency and depth on simulated customer behavior for 100 customers over 53 weeks with all other parameters the same as in the calibration step. (A) (Left) Key metrics averaged across all customers. The average discount plot is showing the discount realized across all transactions. The demand plot is showing the total number of items sold across all weeks. Category penetration is showing the number of categories purchased per customer per week. Customer retention rate is the proportion of customers who make a purchase in the final 4 weeks of the simulation. (Right) Key metrics for customer segments. Customers segmented based on the percentile rank of $\bar{\beta}_{ui} = \sum_{i}{\beta_{ui}^w} / N_i$ with average values of -15.8, -12.8, and -9.5 for the segments with high, medium, and low price sensitivity.  (B) Policy definition based on depth and frequency of discounts across all products and customers. Effective discount is the time-averaged discount rate offered to customers.}  
    \end{minipage}
\end{figure}

To mimic common industry practice, we set design constraints that the discount state should not occur more than 60\% of the time on average and the average discount depth should not exceed 50\%. We then designed scenarios where the time-averaged discount levels ranged from 3\% to 24\% (Table \ref{tab:scenario_disc}). For the lower two discount levels, we ran a low-frequency and high-frequency scenario, where the percentage of time spent in the discount stage was either 30\% or 60\% respectively. To make the pairwise comparison of high- and low-frequency scenarios valid, we kept the time-averaged discount level the same. For the highest discount level, we ran only the high-frequency scenario, so as not to exceed the discount depth constraint.

After running the various scenarios, we computed the aggregate outcome metrics across all customers and products. Intuitively, one expects that the deeper the discount, the larger the impact on the metrics, and this is indeed the case. Looking at Figure \ref{fig:scenario_chart}, we see that the realized discounts (average discounts applied to purchased items) and overall demand are proportional to the discount depth. This implies that customer demand is concentrated during the discounting events, otherwise the realized discounts would be proportional to the effective discount. For policies I and II the effect of discount depth and frequency is muted, indicating a weak response to low discounts. For policies III, IV, and V, we see a sharp trade-off between overall demand and revenue. Here, the selection of an optimum pricing policy depends on whether the firm is prioritizing growth or profitability. Policies IV and V, which offer deeper discounts, resulted in increased demand, higher category penetration, and improved customer retention compared to Policy III, which has shallower discounts. However, these deeper discounts lead to a reduction in revenue due to the pronounced increase in actual discounts used.

These aggregate trends, while significant, mask significant heterogeneity across customers. To assess the potential impact of AI-powered personalization, we split the customer base into segments based on their price sensitivity and analyzed the response across segments. We found that for more price-sensitive customers, the effect of discounts on revenue was muted, which we attribute to these customers having a lower overall propensity to purchase products. For this most price-sensitive segment, there was a slight decrease in revenue moving from III to IV, attributable to the sharp increase in realized discounts. For higher-spending, less price-sensitive customers, the opposite was true, where policy III produced more revenue than IV. So for revenue optimization, it is clear that there is potential uplift from personalizing the discounting policies, e.g. by issuing targeted coupons. For retention and category penetration, there are also differences across segments, with the larger discounts having a more positive impact on the price-sensitive customers which is apparent when comparing policies II with III, IV, and V. These results fit well with the intuition that a retailer can "buy" growth by offering lower prices. The inherent trade-off between growth and profitability underscores the potential of AI, especially RL methods that account for long-term effects, as a powerful tool for retailers. We have shown here that RetailSynth can capture short-term and long-term dynamics and generate differential responses across segments, which gives us the confidence to deploy it for AI systems evaluation tasks.  

\FloatBarrier 

\FloatBarrier

\section{Discussion} \label{discussion}

RetailSynth is designed to generate synthetic retail datasets for evaluating AI systems, which dynamically adjust the marketing mix (price, promotion, product, and place) over time with the aim of maximizing customer lifetime value. As AI systems mature, we expect increasing emphasis on price and promotion, both as decision variables and as key modeling inputs. Given this context, we carefully incorporated price sensitivity into our model so that it systematically varies across customers and products. We also valued scalability, designing RetailSynth to be embarrassingly parallel to run, with each customer and product category being independent. We made many strongly opinionated design decisions, and so we feel it is valuable to compare RetailSynth to existing solutions. To facilitate this comparison, we choose the following evaluation criteria: 
\begin{enumerate}
    \item generating synthetic datasets
    \item modeling the quantity of items purchased per time period
    \item modeling item-to-item interactions i.e. basket and bundling effects, and
    \item modeling price as a dynamic decision variable
\end{enumerate}
RetailSynth fulfills criteria 1, 2, and 4, while neglecting 3, due to the aforementioned scalability concerns. In Table \ref{tab:recent_models_synth}, we show how RetailSynth compares to other recent models and simulation frameworks. Importantly, none of the included models or frameworks meet all the design criteria, which we attribute to the scale and complexity of the modeling problem at hand and also to the evolving objectives of retail AI systems. For classical product recommendation systems or one-time product promotions, the outcome variable of interest is indeed the binary record of product choices per shopping trip, and we see the majority of recent papers focusing on this target variable. It is the change of objective from the more narrow problem of product recommendation or coupon targeting to the more broadly defined customer lifetime value maximization that requires modeling the complete purchase record over time. Our framework can serve as a strong baseline method for building retail AI simulation environments as it captures the relevant outcome variables and models key decision variables. There is, however, room to improve our model's ability to capture item-to-item interactions. We anticipate building on an existing sequential model, such as SHOPPER or directly modeling product bundles \cite{tulabandhula_multi-purchase_2023}. However, as Athey and coworkers \cite{donnelly_counterfactual_2019} noted, modeling item-to-item interactions suffers from the curse of dimensionality and so any such model will require significantly greater computational resources and may be more limited in scale than RetailSynth but more applicable to certain domains. Another area ripe for improvement is accurately capturing the effects of pricing decisions themselves. Our model leverages utility theory to model price sensitivity, but, with a large amount of diverse, training data, a deep learning model could almost certainly learn better the complex multivariate relationships inherent to retail customer decision-making \cite{gabel_product_2022}. 

Looking ahead, our focus is on evolving RetailSynth to simulate a broader range of business conditions. We are especially interested in incorporating dynamic customer preferences to make the modeling process more accurate and useful to retail managers. This would allow simulations to capture any feedback effects from AI-driven marketing mix decisions. As we adapt RetailSynth for additional use cases, we expect to build adapters to existing algorithms and to create use-case specific model configurations introducing the relevant decision variables. Our overarching ambition for RetailSynth is to showcase the broad potential of retail AI systems, inspiring increased experimentation in production settings and community collaboration on the development of simulation environments. 
\begin{table}[ht]
    \centering
    \begin{threeparttable}
        \footnotesize
        \caption{Recent approaches for modeling retail purchase decisions and generating synthetic shopping data.}
        \renewcommand{\arraystretch}{1.5}
        \begin{tabularx}{\linewidth}{L{2.5cm}L{2cm}L{2.5cm}cc}
            \toprule
            \multirow{2}{2.5cm}{\textbf{Model Description}} 
            & \multirow{2}{2cm}{\textbf{Generates Synthetic Shopping Datasets}} 
            & \multirow{2}{\linewidth}{\textbf{Outcome Variables}} 
            & \multicolumn{2}{c}{\textbf{Modeled Effects}} \\
            \cmidrule(r){4-5}
                & & & \parbox[t]{1.7cm}{\centering\textbf{Item-to-Item\\Interactions}} 
                & \parbox[t]{1.7cm}{\centering\textbf{Pricing\\Decisions}} \\
            \addlinespace
            \midrule
            Two-stage nested factorization product choice model \cite{donnelly_counterfactual_2019} & False & Binary - product choices per shopping trip & False & True \\
            \addlinespace
            Three-stage feature-based matrix factorization purchase decision model \cite{wan_modeling_2017} & False & Count - product quantity per shopping trip & False & True \\
            \addlinespace
            SHOPPER: Hierarchical latent variable model of market baskets \cite{ruiz_shopper_2019} & False & Binary - product choices per shopping trip & True & True \\
            \addlinespace
            Deep learning product choice model \cite{gabel_product_2022} & False & Binary - product choices per shopping trip & True & True \\
            \addlinespace
            LSTM-GAN future basket generator \cite{doan_generating_2019} & Partially true - simulates future purchases given customer history & Count - Product quantity purchased per time-period & True & False \\
            \addlinespace
            Two-stage product choice decision synthesizer \cite{gabel_product_2022} & True & Binary - product choices per shopping trip & False & True \\
            \addlinespace
            RetailSynth: Four-stage purchase decision model (this work) & True & Count - Product quantity purchased per time-period & False & True \\
            \addlinespace
            \bottomrule
        \end{tabularx}
        \vspace{0.05cm}
        \label{tab:recent_models_synth}
    \end{threeparttable}
\end{table}

\bibliographystyle{IEEEtran}
\bibliography{retailsynth}

\appendix

\titleformat{\section}[block]
  {\normalfont\Large\bfseries}{\appendixname~\thesection:}{0.5em}{}
\section{Calibrated simulation parameters}\label{appendix:parameter}

\begin{table}[H]
\centering
\begin{threeparttable}
    \small
    \caption{Simulation Parameters}
    \label{tab:simulation_parameters}
    \begin{tabular}{l L{5cm} l}
    
    \toprule
    Parameter & Description & Value \\
    
    \midrule
    
    \multicolumn{3}{l}{\textbf{Store visit choice}} \\
        $\gamma_0^{store}$ & Store base utility & $\gamma_0^{store} \sim \text{Gumbel}(0, 0.1)$ \\
        \addlinespace
        $\gamma_1^{store}$ & Sensitivity to store visit preference score from previous time step & $\gamma_1^{store} \sim TN(0.01, 0.01; 0, \infty)$ \\
        \addlinespace
        $\gamma_2^{store}$ & Sensitivity to customer attributes and marketing effects & $\gamma_2^{store} \sim U(0.2,0.3)$ \\
        \addlinespace
        $\theta$ & Store visit probability decay factor & $\theta \sim U(0.25,0.45)$ \\
        \addlinespace
    
    \midrule
    \addlinespace
    \multicolumn{3}{l}{\textbf{Category choice}} \\
        $\gamma_{0j}^{cate}$ & Category base utility & $\gamma_{0j}^{cate} \sim \mathcal{N}(-5, 0.5)$ \\
        \addlinespace
        $\gamma_{1j}^{cate}$ & Sensitivity to category preference score & $\gamma_{1j}^{cate} \sim TN(0.1, 0.04; 0, 0.12)$ \\
        \addlinespace
    
    \midrule
    \addlinespace
    \multicolumn{3}{l}{\textbf{Product choice}} \\
        $\mathbf{\beta_{ui}^x}$ & Sensitivity to customer and product specific time-varying features & $\mathbf{\beta_{ui}^x} \sim \text{LogNormal}(-1, 1)$ \\
        \addlinespace
        $\beta_{ui}^w$ & Sensitivity to price & $\beta_{ui}^w = c \beta_u^w \beta_i^w$ \\
        & & $c = -1.4$ \\
        & & $\beta_u^w \sim TN(-3, 0.8; -\infty, 0)$ \\
        & & $\beta_i^w \sim TN(-3, 1; -\infty, 0)$ \\
        \addlinespace
        $\beta_{ui}^{z}$ & Sensitivity to features endogenous with price & $\beta_{ui}^{z} \sim \text{LogNormal}(1, 0.4)$ \\
        \addlinespace
        $\epsilon_{uit}$ & Noisy intercept & $\epsilon_{uit} = \text{Gumbel}(0, 0.1)$ \\
        \addlinespace
    
    \midrule
    \addlinespace
    \multicolumn{3}{l}{\textbf{Pricing and discounting}} \\
        $\alpha_{i0}$ & Observed price components & $\alpha_{i0} \sim TN(1.3, 0.5; 0, \infty)$ \\
        \addlinespace
        $\alpha_{1}$ & Sensitivity to unobserved price components & $\alpha_{1} \sim \text{LogNormal}(0.08, 1.2)$ \\
        \addlinespace
        $Z_{i}$ & Unobserved factors influencing price & $Z_{i} \sim \text{TruncatedNormal}(\mu, \sigma)$ \\
        & & $\mu \sim \text{HalfNormal}(0, 1)$ \\
        & & $\sigma \sim \text{HalfNormal}(0, 1)$ \\
        \addlinespace
        
    \midrule
    \addlinespace
    \multicolumn{3}{l}{\textbf{Product demand}} \\
        $\gamma_0^{prod}$ & Baseline product demand & $\gamma_0^{prod} \sim \text{Gumbel}(0, 0.1)$ \\
        \addlinespace
        $\gamma_{ui}^{prod}$ & Sensitivity to product utility & $\gamma_{ui}^{prod} \sim \text{LogNormal}(-4, 0.05)$ \\
        \addlinespace
    
    \midrule
    \addlinespace
    \multicolumn{3}{l}{\textbf{Other factors}} \\
         $\mathbf{X}$ & Customer and product features & $\mathbf{X} \sim \mathcal{N}(\mu, \sigma)$ \\
        & & $\mu \sim \mathcal{N}(0, 1)$ \\
        & & $\sigma \sim \mathcal{N}(0, 1)$ \\
    
    \bottomrule
    \end{tabular}
\end{threeparttable}
\end{table}
\section{Discounting policy definition for scenario analysis}\label{appendix:scenario_def}

The expected discount depth is the mean of the specified Beta distribution ($\alpha_d/(\alpha_d + \beta_d)$). The discount state probability is calculated using the Chapman-Kolmogorov equation given the average transition probabilities, which are also Beta distributed. See Chapter 4 of \cite{ross_introduction_1997} for a thorough description of how to perform this calculation.

\begin{table}[H]
\label{tab:scenario_discount_details}
    \centering
    \begin{threeparttable}
    \footnotesize
    \caption{Policy definition based on depth and frequency of discounts across all products and customers.}
    \begin{tabular}{cp{1.2cm}p{1.2cm}p{1.2cm}ccc}
        \toprule
        \multirow{3}{*}{\textbf{Policy}} &
        \multirow{3}{1.2cm}{\textbf{Effective discount}} &
        \multirow{3}{1.2cm}{\textbf{Discount state \newline probability}} & 
        \multirow{3}{1.2cm}{\textbf{Expected discount depth}}  & 
        \multirow{3}{*}{$\alpha_{01}^{trans},\beta_{01}^{trans}$} & 
        \multirow{3}{*}{$\alpha_{11}^{trans},\beta_{11}^{trans}$} &
        \multirow{3}{*}{$\alpha_d, \beta_d$} \\
        &&&&&&\\
        &&&&&&\\
        \midrule
         I & 3\%   & 60\% & 5\%  & (60,40)&(60,40) & (5,95)\\
         II & 3\%    & 30\% & 10\% & (30,70)& (30,70)& (10,90)\\
         III & 15\%  & 60\% & 25\% &(60,40) & (60,40)&(25,75)\\
         IV & 15\% & 30\% & 50\% &  (30,70)& (30,70) & (50,50)\\
         V & 24\%   & 60\% & 40\% &(60,40) &(60,40) &(40,60)\\
        \bottomrule
    \end{tabular}
    \end{threeparttable}
\end{table}

We reused the coefficients for simulator calibration shown in Table \ref{tab:simulation_parameters} while running scenario analysis, except for the following coefficients to compute the store visit probability.

\begin{table}[H]
\centering
\begin{threeparttable}
    \small
    \caption{Simulation Parameters}
    \label{tab:simulation_parameters_scenario}
    \begin{tabular}{l L{5cm} l}
    
    \toprule
    Parameter & Description & Value \\
    
    \midrule
    
    \multicolumn{3}{l}{\textbf{Store visit choice}} \\
        $\gamma_0^{store}$ & Store base utility & $\gamma_0^{store} \sim \text{Gumbel}(-1.85, 0.1)$ \\
        \addlinespace
        $\gamma_1^{store}$ & Sensitivity to store visit preference score from previous time step & $\gamma_1^{store} \sim TN(0.1, 0.05; 0, \infty)$ \\
        \addlinespace
        $\gamma_2^{store}$ & Sensitivity to customer attributes and marketing effects & $\gamma_2^{store} \sim U(0, 0.001)$ \\
        \addlinespace
        $\theta$ & Store visit probability decay factor & $\theta = 0$ \\
        \addlinespace
    \bottomrule
    \end{tabular}
\end{threeparttable}
\end{table}

\section{Price elasticity for synthetic data model}\label{appendix:elasticity}

We provide detailed calculations of the price elasticities at each stage of our multi-stage retail decision-making model. We follow the approach outlined in \cite{wan_modeling_2017} to calculate the elasticity for our model specification. 

Recall from Section \ref{sec:model_spec} that the probability of customer $u$ purchasing $Q$ quantity of product $i$ can be expressed using the conditional probability:

\begin{align*}
&P(Q_{ui}=q,S_u,C_{uj},B_{ui})\\
&=P(S_u)P(C_{uj} | S_u) P(B_{ui}|S_u,C_{uj})P(Q_{ui}=q | S_u,C_{uj},B_{ui})  
\end{align*}

where:

\begin{align*}
S_u &= \text{binary outcome (visited versus did not visit store) for customer $u$} \\
C_{uj} &= \text{binary outcome (category purchased versus did not purchase) } \\
          &\text{\quad$\>$ for customer $u$ and category $j$}\\
B_{ui} &= \text{binary outcome (product purchased versus did not purchase)} \\
          &\text{\quad$\>$ for customer $u$ and product $i$}\\
Q_{ui} &= \text{units purchased by customer $u$ of product $i$}\\
\end{align*}

The expected quantity purchased can be expressed as:
\begin{align*}
    &\mathbb{E}(Q_{ui}=q,S_u,C_{uj},B_{ui})\\
    &=\mathbb{E}(S_u)\mathbb{E}(C_{uj} | S_u) \mathbb{E}(B_{ui}|S_u,C_{uj})\mathbb{E}(Q_{ui}=q | S_u,C_{uj},B_{ui})
\end{align*}

We define the overall elasticity of the system at time $t$ for customer $u$ and product $i$ as the proportional change in expected quantity purchased relative to the proportional change in price:

\begin{align}
    e_{uit}^{overall} &= \left. \frac{d \mathbb{E}Q_{uit}}{\mathbb{E}Q_{uit}} \middle/ \frac{d P_{uit}}{P_{uit}} \right.\\
    &= e_{uit}^{store} + e_{uit}^{cate} + e_{uit}^{prod} + e_{uit}^{quant}
\label{eq:overall_elasticity}
\end{align}

The independence assumption we impose on the different choice stages of the framework yields an elegant form of elasticity that is additive in nature across the various choice stages. In the following subsections, we will derive the elasticity for the individual decision stages.

\subsection{Store visit}
The store visit decision has no dependence on current prices in the calibrated model. (The Complete Journey data covers a cohort specifically selected to have high loyalty and repeat purchasing behavior, so we used a simplified form of the model.) 

For the scenario analysis, we wanted to include the effect of store marketing on the retention of a heterogeneous customer base and so as described in Equation \ref{eq:store_utility_scenario}, we included the discount level $D_{it}$ in the store utility equation. For this variant of the model, the elasticity is 

\begin{align}
    e^{store}_{uit} &= \left. \frac{d p^{store}_{ut}} {p^{store}_{ut}} \middle/ \frac{d P_{uit}}{P_{uit}} \right.\\
    &= -\gamma_2(1-p^{store}_{ut})(1-D_{it})
\end{align}

\subsection{Category choice}
We can derive the category elasticity for customer $u$ and product $i$ to be
\begin{align}
    e^{cate}_{uit} &= \left. \frac{d p^{cate}_{ujt}} {p^{cate}_{ujt}} \middle/ \frac{d P_{uit}}{P_{uit}} \right.\\
    &= p^{prod}_{uit} (1 - p^{cate}_{ujt})
    \gamma^{cate}_{1j}\beta_{ui}^{w}
\end{align}

\subsection{Product choice}
With this choice model, we can derive the product self elasticity for each product as

\begin{align}
    e^{prod}_{u(ii)t} &= \left. \frac{d p^{prod}_{uit}} {p^{prod}_{uit}} \middle/ \frac{d P_{uit}}{P_{uit}} \right.\\
    &= (1 - p^{prod}_{uit}) \beta_{ui}^{w}
\end{align}

\subsection{Product demand}
The demand elasticity is derived as follows
\begin{align}
    e^{quant}_{u(ii)t} &= \left. \frac{d (\lambda_{uit} + 1)} {\lambda_{uit} + 1} \middle/ \frac{d P_{uit}}{P_{uit}} \right.\\
    &= (\frac{\lambda_{uit}}{1 + \lambda_{uit}})\gamma_{ui}^{prod}\beta_{ui}^{w}
\end{align}

\end{document}